# A Mathematical Model for Two Solutes Transport in a Poroelastic Material and Its Applications

Roman Cherniha [a,b1], Joanna Stachowska-Pietka [c], and Jacek Waniewski [c]

[a] *Institute of Mathematics, NAS of Ukraine, 3, Tereshchenkivs'ka Street, 01601 Kyiv, Ukraine*
[b] *School of Mathematical Sciences, University of Nottingham,*
*University Park, Nottingham NG7 2RD, UK*
[c] *Nalecz Institute of Biocybernetics and Biomedical Engineering, PAS,*
*Ks. Trojdena 4, 02 796 Warsaw, Poland*

**Abstract**

Using well-known mathematical foundations of the elasticity theory, a mathematical model for two solutes transport in a poroelastic material (soft tissue is a typical example) is suggested. It is assumed that molecules of essentially different sizes dissolved in fluid and are transported through pores of different sizes. The stress tensor, the main force leading to the material deformation, is taken not only in the standard linear form but also with an additional nonlinear part. The model is constructed in 1D space and consists of five nonlinear equations. It is shown that the governing equations are integrable in stationary case, therefore all steady-state solutions are constructed. The obtained solutions are used in an example for healthy and tumour tissue, in particular, tissue displacements are calculated and compared for parameters taken from experimental data in cases of the linear and nonlinear stress tensors. Since the governing equations are non-integrable in non-stationary case, the Lie symmetry analysis is used in order to construct time-dependent exact solutions. Depending on parameters arising in the governing equations, several special cases with non-trivial Lie symmetries are identified. As a result, multi-parameter families of exact solutions are constructed including those in terms of special functions(hypergeometric and Bessel functions). A possible application of the solutions obtained is demonstrated.

# 1 Introduction

The basic theory of poroelasticity was developed by Biot [1]. The modern poroelastic theory was formulated by [2–4] and soon applied in geology for the description of penetration of water or oil across soil or cracked rock [2], [5–7]. Later on, the theory found its applications in biology, medicine and biomedical engineering for the description of different anatomical structures as vertebrate disc, cornea and permselective membranes [8–10], [11–13], [14], [4,15], [16–18]. Its specific extension includes internal sources inside the tissue as the capillary and lymphatic beds in the muscle [4, 8–10, 13, 14]. However, most biological models neglect the elastic characteristics of the tissue and focus on the transport of fluid and solute inside and out of the tissue, as in peritoneal dialysis [19], whereas for the non-perfused tissues the poroelastic description may include both phenomena: transport processes and elastic changes of the tissue volume [4, 15, 20–22]. Beside the hydrostatic pressure of fluid also the osmotic pressure of dissolved solutes may be of importance, especially in biological applications [9, 14, 21–24]. The poroelastic theory considers the system formed by a solid, elastic matrix with pores that can be penetrated by a fluid and dissolved solutes. The deformation of the system under the fluid pressure is described by a deformation vector, and the dynamics of the deformation under the forces needs in general to be described by second order tensors [2–4]. The relationship between stress and strain is usually assumed to be linear, but we discuss a simple generalization to nonlinear relationship. The flux of fluid depends on the hydrostatic pressure and solute concentration gradients (osmotic pressure), whereas the fluxes of solutes include diffusive and convective transport mechanisms [9, 12, 21, 22, 27].

The general three-dimensional theory of poroelasticity with fluid transport is mathematically complex even in symmetric systems because relevant mathematical models involve 3D nonlinear partial differential equations (PDEs). Moreover if one takes into account time-dependence then (1+3)-dimensional PDEs should be used. A general mathematical description of this theory can be found in the excellent book [28] (see also references therein). However, if fluid transport is replaced by the solute transport involving molecules of different sizes then the model is more complicated. To the best of our knowledge, there are no papers presenting rigorous mathematical models for solute transport in poroelastic materials in the general case. As a result, at least for biomedical applications, the one-dimensional version was only discussed [29]. Moreover, experimental and clinical data on solute and fluid penetration into / removal from the tissue are mostly presented as 1D problem, i.e. the depth into tissue from the surface (see for instance [14] and the relevant references therein).

Frequently the applications of the theory include ad hoc assumptions that allow to simplify the relevant mathematical models in order to get analytical results [17, 18, 26]. Here we used several assumptions in order to formulate a non-stationary mathematical model for two solutes transport in a poroelastic material in the 1D space dimension. The classical Lie method [31–35] was applied in order to find exact solutions of the model obtained. This method is one of the most powerful tools for constructing exact solutions of nonlinear PDEs and nowadays is widely used for solving real-world models. For instance, there are several recent studies devoted to search for Lie symmetry of the PDE systems arising in life sciences [36]– [41].

The remainder of the paper is organized as follows. Section 2 is devoted to deriving the afore-mentioned model. The resulting model consists of six PDEs of parabolic and hyperbolic types. These governing equations should be supplied by the relevant boundary conditions in order to get real-world models. In Section 3, steady-state solutions of the model are constructed and discussed. We illustrate the obtained results by an example for healthy and tumour tissue.

In particular, impact of the nonlinear term in the stress tensor on the displacement is analyzed. Section 4 presents Lie symmetries of the governing equations of the model. It is shown that different Lie symmetry operators occur depending on the model parameters. Section 5 is devoted to the construction of time-dependent solutions. Firstly (Subsection 5.1) we give some basic information about the technique used, furthermore the technique is applied for search for exact solutions. Finally, we discuss the main results obtained in the last section.

## 2 Derivation of mathematical model in 1D approximation

The mathematical model for the poroelastic materials (PEM) with the variable volume is developed under the following assumptions:

1. 1D approximation in space;

2. PEM consists of pores of two different sizes ('small' and 'large') and matrix;

3. no internal sources/sinks (however, they may be added);

4. incompressible fluid;

5. molecules of essentially different sizes dissolved in fluid are transported through pores;

6. isothermal conditions for transport in PEM.

The governing equations of the model consists of continuity equations. The first one expresses the volume balance for an infinitesimal volume element $dV$ of PEM as:

$$\frac{1}{dV}\frac{\partial dV}{\partial t} = \frac{\partial e}{\partial t} = \frac{\partial}{\partial t}\left(\frac{\partial u}{\partial x}\right) = -\frac{\partial}{\partial x}\bar{j}_V \tag{1}$$

where the deformation vector $u$ is defined as

$$u(x, t; x_0) = x(t) - x_0, \ x_0 = x(0), \tag{2}$$

the function $e = \frac{\partial u}{\partial x}$ is called dilatation and $\bar{j}_V$ is the volumetric flux across the PEM. For the derivation of this equation and subsequent those we used the theory described in [3, 4, 6]. It comes from continuity law for the mass $\rho dV$ of the infinitesimal element $dV$ and the mass density $\rho$ that:

$$\frac{1}{dV}\frac{\partial(\rho dV)}{\partial t} = \frac{\partial \rho}{\partial t} + \rho\frac{\partial e}{\partial t} = -\frac{\partial}{\partial x}\bar{j}_\rho$$

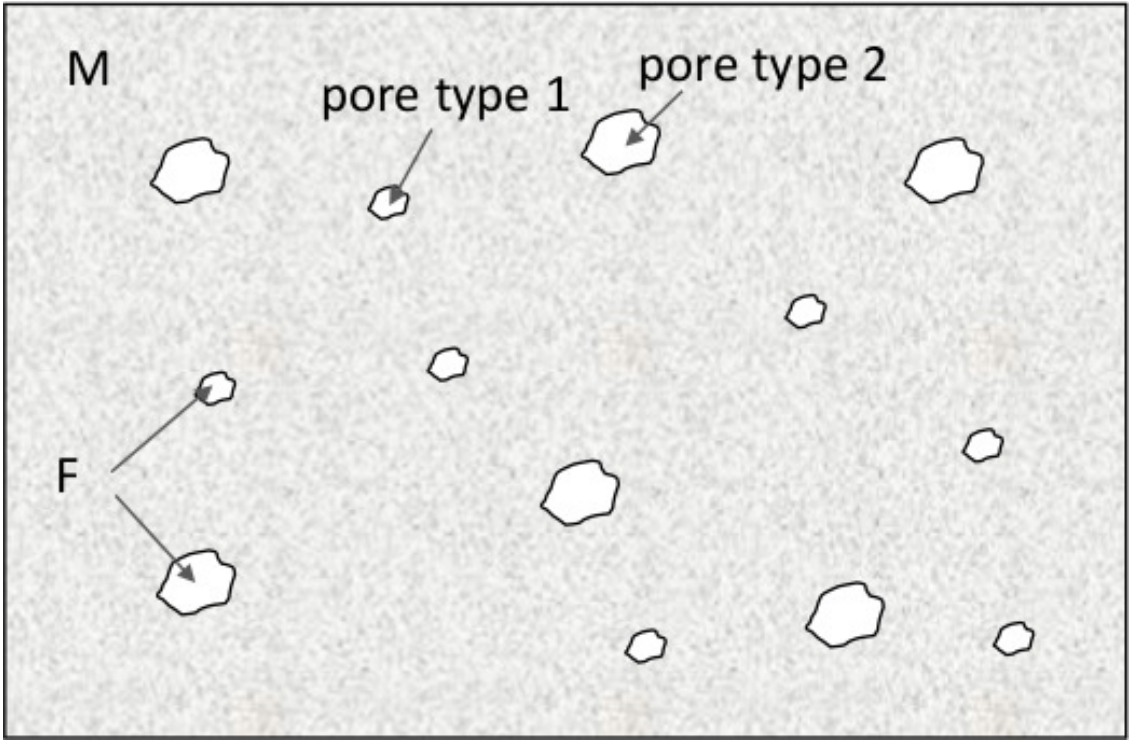

Figure 1: Cross-section of the PEM layer (perpendicular to the axis x) with schematic representation of the structural components of the PEM layer trough which the fluid and solute transport occurs. $M$ denotes the matrix, $F$ is the fluid phase consisting of small pores 1 and large pores 2.

i.e.
$$\frac{\partial \rho}{\partial t} = -\frac{\partial}{\partial x}\bar{j}_\rho + \rho\frac{\partial}{\partial x}\bar{j}_V, \tag{3}$$

where $\bar{j}_\rho$ is the flux of mass with density $\rho$ through the PEM.

Let us consider PEM with two phases distinguished: fluid phase, F, with the fractional volume $\theta_F$ and solid phase i.e. matrix phase, M, with the fractional volume $\theta_M$ (see Fig.1). Because PEM consists of pores of two different sizes, we may write down the relations

$$\theta_1 + \theta_2 + \theta_M = \theta_F + \theta_M = 1, \tag{4}$$

where $\theta_1$ and $\theta_2$ stand for the fractional volume of small pores and large those, respectively. Assuming that a solute, which consists of molecules of essentially different sizes ('small' and 'large') is transported, we realize that small molecules (e.g. glucose) are able to go through the both pores, i.e. the fractional volume for their transport is $\theta_1 + \theta_2 = \theta_F$, while large molecules (e.g. albumin) can be transported only through the large pores, i.e. the relevant fractional volume is $\theta_2$. In what follows we assume $\theta_2 = \alpha\theta_F$, $0 \leq \alpha < 1$ for simplicity.

**Remark 1.** In the general case, the relation between $\theta_2$ and $\theta_F$ is unknown, hence one should to add an additional equation for $\theta_2$. The latter can be written down in a similar way as it was done in [25].

So, the volume balance for each phase separately is:

$$\frac{1}{dV}\frac{\partial (\theta_F dV)}{\partial t} = \frac{\partial \theta_F}{\partial t} + \theta_F \frac{\partial e}{\partial t} = -\frac{\partial}{\partial x}\bar{j}_{VF} \tag{5}$$

$$\frac{1}{dV}\frac{\partial (\theta_M dV)}{\partial t} = \frac{\partial \theta_M}{\partial t} + \theta_M \frac{\partial e}{\partial t} = -\frac{\partial}{\partial x}\bar{j}_{VM} \tag{6}$$

so that

$$\frac{\partial \theta_F}{\partial t} = -\frac{\partial}{\partial x}\bar{j}_{VF} + \theta_F \frac{\partial}{\partial x}\bar{j}_V \tag{7}$$

$$\frac{\partial \theta_M}{\partial t} = -\frac{\partial}{\partial x}\bar{j}_{VM} + \theta_M \frac{\partial}{\partial x}\bar{j}_V, \tag{8}$$

where $\bar{j}_{VF}$ and $\bar{j}_{VM}$ corresponds to the volumetric fluxes in fluid and matrix phases, respectively. Similarly for the densities the equations

$$\frac{1}{dV}\frac{\partial (\rho_F \theta_F dV)}{\partial t} = \frac{\partial (\rho_F \theta_F)}{\partial t} + \rho_F \theta_F \frac{\partial e}{\partial t} = -\frac{\partial}{\partial x}(\rho_F \bar{j}_{VF}) \tag{9}$$

$$\frac{1}{dV}\frac{\partial (\rho_M \theta_M dV)}{\partial t} = \frac{\partial (\rho_M \theta_M)}{\partial t} + \rho_M \theta_M \frac{\partial e}{\partial t} = -\frac{\partial}{\partial x}(\rho_M \bar{j}_{VM}) \tag{10}$$

take place, with densities $\rho_F$ and $\rho_M$ in fluid and matrix phases, respectively.

Note that the relations

$$\rho = \rho_1 \theta_1 + \rho_2 \theta_2 + \rho_M \theta_M = \rho_F \theta_F + \rho_M \theta_M \tag{11}$$

holds, where $\rho_1$ stands for the fluid density involving only small molecules, while $\rho_2$ stands for the fluid density involving both types of molecules. Taking into account the above assumption $\theta_2 = \alpha\theta_F$, equation (11) leads to the relation

$$\rho_F = (1-\alpha)\rho_1 + \alpha\rho_2. \tag{12}$$

In what follows we assume that fluid is incompressible, hence $\rho_F = \rho_F^0$ is constant.

Both types of molecules are dissolved in the fluid. Let us denote as $c_1(t,x)$ and $c_2(t,x)$ the concentrations of small and large molecules, respectively. Now we might write down the corresponding equations for the both concentrations:

$$\frac{\partial (\theta_F c_1)}{\partial t} + \theta_F c_1 \frac{\partial e}{\partial t} = -\frac{\partial}{\partial x}\bar{j}_1, \tag{13}$$

$$\frac{\partial (\alpha \theta_F c_2)}{\partial t} + \alpha \theta_F c_2 \frac{\partial e}{\partial t} = -\frac{\partial}{\partial x}\bar{j}_2, \tag{14}$$

where $\bar{j}_1$ and $\bar{j}_2$ are the solute fluxes across the fractional volumes $\theta_F$ and $\theta_2 = \alpha \theta_F$, respectively.

Finally, the dynamics of the PEM element of the mass $\rho dV$ with the velocity $\frac{\partial x}{\partial t} = \frac{\partial u}{\partial t}$ is described by the Newton law:

$$\frac{1}{dV}\frac{\partial \left(\rho \frac{\partial u}{\partial t} dV\right)}{\partial t} = \rho \frac{\partial^2 u}{\partial t^2} + \frac{\partial u}{\partial t}\left(\frac{\partial \rho}{\partial t} + \rho \frac{\partial e}{\partial t}\right) = \frac{\partial}{\partial x}\tilde{\tau}_t \tag{15}$$

where $\tilde{\tau}_t$ is the Terzaghi effective stress tensor [42]. This tensor in the linear poroelastic theory for isotropic materials in 1D approximation can be defined as

$$\tilde{\tau}_t = (\lambda + 2\mu)\frac{\partial u}{\partial x} - p + \gamma_0 RT c_1 + (\gamma_1 RT c_1 + \gamma_2 RT c_2), \tag{16}$$

where $\lambda + 2\mu$ is the elastic modulus with Lame constants $\lambda$ and $\mu$, and the effective pressure is given by the difference between hydrostatic pressure $p$ and osmotic pressures created by the solute concentrations in large and small pores. Hereinafter the constant $RT$ is gas constant times temperature, which is also some constant. In small pores, the osmotic pressure is equal to $\gamma_0 RT c_1$ with a coefficient $\gamma_0 \leq 1$, which takes into account the size of small molecules and a characteristic size of pores (e.g. the average diameter of small pores). In large pores, the osmotic pressure is equal to $\gamma_1 RT c_1 + \gamma_2 RT c_2$. Here the coefficients $\gamma_1 \leq 1$ and $\gamma_2 \leq 1$ take into account the sizes of small and large molecules in large pores, respectively. A reasonable choice can be $\gamma_0 + \gamma_1 = \sigma_1$, $\gamma_2 = \alpha \sigma_2$, where $\sigma_1$ and $\sigma_2$ are reflection coefficients of PEM for solutes with small and large molecules, respectively.

Generally speaking, this tensor can depend nonlinearly on the dilatation $e = \frac{\partial u}{\partial x}$. In this case, we obtain

$$\tilde{\tau}_t = -p + (\gamma_0 + \gamma_1)RT c_1 + \gamma_2 RT c_2 + (\lambda + 2\mu)\frac{\partial u}{\partial x} + \kappa\left(\frac{\partial u}{\partial x}\right)^2, \tag{17}$$

where $\kappa$ can be thought as a parameter. In the case, when the coefficients $\gamma_i \leq 1$, $i = 0, 1, 2$ are specified as above, the nonlinear tensor (17) takes the form

$$\tilde{\tau}_t = -p + \sigma_1 RT c_1 + \alpha \sigma_2 RT c_2 + (\lambda + 2\mu)\frac{\partial u}{\partial x} + \kappa\left(\frac{\partial u}{\partial x}\right)^2. \tag{18}$$

Now one needs to specify the fluxes across the PEM. In the fluid phase, we have

$$\bar{j}_{VF} = -k\Big(\frac{\partial p}{\partial x} - \sigma_1 RT\frac{\partial c_1}{\partial x} - \alpha\sigma_2 RT\frac{\partial c_2}{\partial x}\Big) + \theta_F\frac{\partial u}{\partial t} \qquad (19)$$

where the first term means the volumetric fluid flux relative to the matrix calculated according to the extended Darcy's low with hydraulic conductivity $k$. Hereinafter the coefficient $\alpha$ reflects the fact that only a part, $\alpha\theta_F$, of the fractional fluid volume is accessible for large molecules.

Volumetric matrix flux is given by the equation

$$\bar{j}_{VM} = \theta_M\frac{\partial u}{\partial t} = (1-\theta_F)\frac{\partial u}{\partial t} \qquad (20)$$

In contrast to the fluxes defined above, the solute fluxes through the PEM includes diffusive and convective terms. The solute flux created by diffusion and convection of small molecules has the form

$$\bar{j}_1 = -D_1\frac{\partial c_1}{\partial x} + S_1 c_1\Big(\bar{j}_{VF} - \theta_F\frac{\partial u}{\partial t}\Big) + \theta_F c_1\frac{\partial u}{\partial t}, \qquad (21)$$

where $D_1$ stands for solute diffusivity in PEM, and $S_1$ is the sieving coefficient of solute with small molecules. Similarly, the solute flux created by diffusion and convection of large molecules reads as

$$\bar{j}_2 = -\alpha D_2\frac{\partial c_2}{\partial x} + \alpha S_2 c_2\Big(\bar{j}_{VF} - \theta_F\frac{\partial u}{\partial t}\Big) + \alpha\theta_F c_2\frac{\partial u}{\partial t}, \qquad (22)$$

where $D_2$ is the solute diffusivity and $S_2$ is the sieving coefficient of solute with large molecules.

Finally, the fluxes arising in equations (1) and (3) must be related with the above defined fluxes. So, the relations

$$\bar{j}_V = \bar{j}_{VF} + \bar{j}_{VM} = -k\Big(\frac{\partial p}{\partial x} - \sigma_1 RT\frac{\partial c_1}{\partial x} - \alpha\sigma_2 RT\frac{\partial c_2}{\partial x}\Big) + \frac{\partial u}{\partial t} \qquad (23)$$

and

$$\bar{j}_\rho = \rho_F\bar{j}_{VF} + \rho_M\bar{j}_{VM} = -k\Big(\frac{\partial p}{\partial x} - \sigma_1 RT\frac{\partial c_1}{\partial x} - \alpha\sigma_2 RT\frac{\partial c_2}{\partial x}\Big) + \rho\frac{\partial u}{\partial t} \qquad (24)$$

should take place.

Thus, equations (7)–(10) and (13)–(15) form a system of seven partial differential equations for seven variables $\theta_F$, $\theta_M$, $\rho_M$, $u$, $c_1$, $c_2$, and $p$. However, there are the known relations for the two phase PEM (4), so that the number of governing equations can be reduced to six. In fact, if we consider the fluid, say water, as an incompressible fluid then $\rho_F = \rho_F^0 = const$. Substituting the later into (9) and taking into account (1), one obtains the equation

$$\rho_F^0\frac{\partial \theta_F}{\partial t} - \rho_F^0\theta_F\frac{\partial}{\partial x}\bar{j}_V = -\rho_F^0\frac{\partial}{\partial x}\bar{j}_{VF}.$$

Obviously, the above equation is equivalent to (7), therefore only three equations among (7)–(10) are independent. In particular, one may take (7), (8) and the third equation in the form of the sum of the inequivalent equations (9) and (10):

$$\frac{\partial \rho}{\partial t} - \rho \frac{\partial}{\partial x} \bar{j}_V = -\frac{\partial}{\partial x} \left( \rho_F \bar{j}_{VF} + \rho_M \bar{j}_{VM} \right). \tag{25}$$

As a result, a six-component system of PDEs to find the unknown functions $\theta_F$, $\rho$, $u$, $c_1$, $c_2$ and $p$ is obtained. The remaining functions $\theta_M$ and $\rho_M$ are calculated using the algebraic expressions (4) and (11).

All the notations arising in the above formulae are explained in Table 1.

Table 1: Description of the symbols used in the formulae above.

| Symbol | Description |
|---|---|
| $dV$ | an infinitesimal volume element of PEM |
| $u$ | deformation vector |
| $e$ | dilatation |
| $\bar{j}_V$ | volumetric flux across the PEM |
| $\rho$ | mass density |
| $j_\rho$ | mass flux across the PEM |
| $\theta_F$ | fractional volume of fluid phase $F$ |
| $\theta_M$ | fractional volume of matrix phase $M$ |
| $\bar{j}_{VF}$ | fluid flux in phase $F$ |
| $\bar{j}_{VM}$ | fluid flux in phase $M$ |
| $\rho_F$ | mass density of fluid phase $F$ |
| $\rho_M$ | mass density of matrix phase $M$ |
| $c_1$, $c_2$ | concentrations of small and large molecules in fluid |
| $\bar{j}_1$, $\bar{j}_2$ | fluxes across the PEM of small and large molecules |
| $\tilde{\tau}_t$ | Terzaghi effective stress tensor |
| $p$ | hydrostatic pressure in PEM |
| $RT$ | gas constant times temperature |
| $\lambda + 2\mu$ | elastic modulus with Lame constants $\lambda$ and $\mu$ |
| $k$ | hydraulic conductivity |
| $D_1$, $D_2$ | diffusivities of small and large molecules |
| $S_1$, $S_2$ | sieving coefficients of small and large molecules |
| $\sigma_1$, $\sigma_2$ | reflection coefficients of small and large molecules |

Now we write down the governing equations in an explicit form. Let us substitute the expressions for equations (19)–(24) into equations (7)–(8), (13)–(14)and (25), and the Terza-ghi effective stress tensor into (15). Making relevant calculations, we arrive at the following nonlinear system consisting of six PDEs for finding unknown functions $\theta_F$, $\rho$, $u$, $c_1$, $c_2$, and $p$:

$$2u_{tx} = k\Big(p_{xx} - T_1 c_{1,xx} - \alpha T_2 c_{2,xx}\Big), \tag{26}$$

$$\rho_t + \rho_x u_t = k(\rho_F^0 - \rho)\Big(p_{xx} - T_1 c_{1,xx} - \alpha T_2 c_{2,xx}\Big), \tag{27}$$

$$\theta_{Ft} + \theta_{Fx} u_t = k(1 - \theta_F)\Big(p_{xx} - T_1 c_{1,xx} - \alpha T_2 c_{2,xx}\Big), \tag{28}$$

$$(\theta_F c_1)_t + (\theta_F c_1)_x u_t + 2\theta_F c_1 u_{tx} = D_1 c_{1,xx} + k S_1 \Big(c_1(p_x - T_1 c_{1,x} - \alpha T_2 c_{2,x})\Big)_x, \tag{29}$$

$$(\theta_F c_2)_t + (\theta_F c_2)_x u_t + 2\theta_F c_2 u_{tx} = D_2 c_{2,xx} + k S_2 \Big(c_2(p_x - T_1 c_{1,x} - \alpha T_2 c_{2,x})\Big)_x, \tag{30}$$

$$\rho u_{tt} + \rho_t u_t + \rho u_t u_{tx} = (\lambda + 2\mu) u_{xx} + \frac{dF}{du_x} u_{xx} - \Big(p_x - (\gamma_0 + \gamma_1) RT c_{1,x} - \gamma_2 RT c_{2,x}\Big), \tag{31}$$

where $T_1 = \sigma_1 RT$ and $T_2 = \sigma_2 RT$ and the lower subscripts $t$ and $x$ denote differentiation with respect to these variables.

By the application of the substitution

$$p_x^* = p_x - T_1 c_{1,x} - \alpha T_2 c_{2,x} \tag{32}$$

where the function $p^*$ corresponds to the so-called effective pressure, the above system of PDEs takes the simpler form:

$$2u_{tx} = k p_{xx}^*, \tag{33}$$

$$\rho_t + \rho_x u_t = k(\rho_F^0 - \rho) p_{xx}^* \tag{34}$$

$$\theta_{Ft} + \theta_{Fx} u_t = k(1 - \theta_F) p_{xx}^*, \tag{35}$$

$$(\theta_F c_1)_t + (\theta_F c_1)_x u_t + 2\theta_F c_1 u_{tx} = D_1 c_{1,xx} + kS_1(c_1 p^*_x)_x, \tag{36}$$

$$(\theta_F c_2)_t + (\theta_F c_2)_x u_t + 2\theta_F c_2 u_{tx} = D_2 c_{2,xx} + kS_2(c_2 p^*_x)_x, \tag{37}$$

$$\rho u_{tt} + \rho_t u_t + \rho u_t u_{tx} = \lambda^* u_{xx} + 2\kappa u_x u_{xx} - \left(p^*_x - (\gamma_0+\gamma_1-\sigma_1)RTc_{1,x} - (\gamma_2-\alpha\sigma_2)RTc_{2,x}\right), \tag{38}$$

where $u(t,x)$, $\rho(t,x)$, $p^*(t,x)$, $\theta_F(t,x)$, $c_1(t,x)$, and $c_2(t,x)$ are to-be-determined functions, while $k > 0$, $\lambda^* = \lambda + 2\mu > 0, 0 \leq S_i \leq 1$, and $D_i \geq 0$ ($i = 1, 2$) are known constants.

Notably, the last equation in the above system reduces to the form

$$\rho u_{tt} + \rho_t u_t + \rho u_t u_{tx} = \lambda^* u_{xx} + 2\kappa u_x u_{xx} - p^*_x \tag{39}$$

by choosing $\gamma_0 + \gamma_1 = \sigma_1$, $\gamma_2 = \alpha\sigma_2$.

In order to complete the mathematical model for the two-phase PEM with variable volume, one needs to add the corresponding boundary conditions and initial profiles. Examples of such conditions are given in Section 5.

## 3  Steady-state solutions of the model

Here we find steady-state solutions of the PDE system (33)–(38) using the standard *ansatz* (this German word is commonly used for substitutions reducing PDEs to ODEs)

$$\begin{aligned} & u = U(x), \ \rho = \mathcal{R}(x), \ p^* = P^*(x), \\ & \theta_F = \Theta_F(x), \ c_1 = C_1(x), \ c_2 = C_2(x). \end{aligned} \tag{40}$$

Substituting ansatz (40) into the system, one obtains the reduced system of ODEs

$$\begin{aligned} & P^{*\prime\prime} = 0, \ D_1 C_1'' + kS_1 P^{*\prime} C_1' = 0, \ D_2 C_2'' + kS_2 P^{*\prime} C_2' = 0, \\ & \lambda^* U'' + \tfrac{dF}{dU'} U'' - \left(P^{*\prime} - (\gamma_0+\gamma_1-\sigma_1)RTC_1' - (\gamma_2-\alpha\sigma_2)RTC_2'\right) = 0. \end{aligned} \tag{41}$$

Hereinafter primes mean the differentiation w.r.t. the variable $x$. Note that two equations, (34)–(35) are automatically fulfilled because $P^{*\prime\prime} = 0$.

The first three equations in (41) do not involve the function $U(x)$ and can be easily integrated. As a result, one obtains

$$\mathcal{R} = \mathcal{R}_0(x), \ \Theta_F = \Theta_{0,F}(x), \ P^* = P_0 + P_1 x,$$

$$C_1 = \begin{cases} A_{01} + A_1 \exp\left(-\frac{kS_1P_1}{D_1}x\right), & S_1 D_1 P_1 \neq 0, \\ A_{01} + A_1 x, & S_1 P_1 = 0, D_1 \neq 0, \\ C_1(x), & S_1 P_1 = D_1 = 0, \end{cases} \quad (42)$$

$$C_2 = \begin{cases} A_{02} + A_2 \exp\left(-\frac{kS_2P_1}{D_2}x\right), & P_1 S_2 D_2 \neq 0, \\ A_{02} + A_2 x, & S_2 P_1 = 0, D_2 \neq 0, \\ C_2(x), & S_2 P_1 = D_2 = 0, \end{cases}$$

where $P_1 \neq 0$, $P_0$, $A_i$, $A_{0i}$ are arbitrary constants, $\mathcal{R}_0(x)$, $\Theta_{0,F}(x)$ and $C_i(x)$ ($i = 0, 1$) are arbitrary smooth functions. In what follows we restrict ourselves to the most realistic case when $S_i D_i \neq 0$, $i = 1, 2$.

**Remark 2.** Steady-state solutions for the hydrostatic pressure $p$ can be easily expressed using formula (32):

$$P = P_0 + P_1 x + T_1 C_1(x) + T_2 C_2(x).$$

Obviously, the last ODE in (41) is reducible to a first-order ODE by straightforward integration, hence one obtains

$$\lambda^* U' + F(U') = U_1 + P_1 x - (\gamma_0 + \gamma_1 - \sigma_1)RT C_1 - (\gamma_2 - \alpha \sigma_2)RT C_2. \quad (43)$$

In the case when $F$ is a linear function, we may set $F = 0$ without losing a generality. In other words, one considers the linear tensor (16). Thus, the following deformation is derived:

$$U = U_0 + \frac{U_1}{\lambda^*}x + U_2 x^2 + U_3 \exp\left(-\frac{kS_1 P_1}{D_1}x\right) + U_4 \exp\left(-\frac{kS_2 P_1}{D_2}x\right) \quad (44)$$

where $U_0$ and $U_1$ are arbitrary parameters, $U_2 = \frac{P_1}{2\lambda^*}$, $U_3 = \frac{(\gamma_0 + \gamma_1 - \sigma_1)RT A_1 D_1}{\lambda^* k S_1 P_1}$, and $U_4 = \frac{(\gamma_2 - \alpha \sigma_2)RT A_2 D_2}{\lambda^* k S_2 P_1}$.

Now we consider the case when a nonlinear poroelastic theory is applied with the tensor (17), i.e.

$$\tilde{\tau}_t = -p + (\gamma_0 + \gamma_1)RT c_1 + \gamma_2 RT c_2 + (\lambda + 2\mu) u_x + \kappa u_x^2, \quad (45)$$

where $\kappa$ is a parameter. In this case, ODE (43) takes the form

$$\lambda^* U' + \kappa (U')^2 = U_1 + P_1 x - (\gamma_0 + \gamma_1 - \sigma_1)RT C_1 - (\gamma_2 - \alpha \sigma_2)RT C_2 \equiv G(x) \quad (46)$$

and is solvable w.r.t. the variable $U'$:

$$U' = \frac{\lambda^*}{2\kappa}\left(-1 \pm \sqrt{1 + \frac{4\kappa}{(\lambda^*)^2}G(x)}\right). \quad (47)$$

Thus, the general solution of ODE (46) is

$$U = \frac{\lambda^*}{2\kappa}\Big(-x \pm \int_{x_0}^{x}\sqrt{1 + \frac{4\kappa}{(\lambda^*)^2}G(x)}dx\Big). \tag{48}$$

Taking into account the functions for $C_1$ and $C_2$ from (42), one realizes that the integral in (48) cannot be expressed via elementary functions provided $S_iD_i \neq 0$, $i = 1, 2$. However, we may assume that the function $|\frac{4\kappa}{(\lambda^*)^2}G(x)|$ is sufficiently small provided $\frac{|\kappa|}{(\lambda^*)^2}$ is sufficiently small. So, using the Taylor series, one obtains

$$\sqrt{1 + \frac{4\kappa}{(\lambda^*)^2}G(x)} = 1 + \frac{1}{2}\frac{4\kappa}{(\lambda^*)^2}G(x) + \frac{1}{8}\Big(\frac{4\kappa}{(\lambda^*)^2}G(x)\Big)^2 + \ldots \tag{49}$$

Thus, substituting (49) into (48), taking the upper sign "+" (otherwise the result will be unrealistic) and making simple calculations, we arrive at the approximate solution

$$U \approx \frac{1}{\lambda^*}\int_{x_0}^{x}G(x)dx + \frac{\kappa}{(\lambda^*)^3}\int_{x_0}^{x}\Big(G(x)\Big)^2 dx. \tag{50}$$

Note that integrals in (50) can be easily expressed via elementary functions.

Now we realize that the approximate solution (50) with $\kappa = 0$ coincides with the exact solution (44). The second term in (50) shows the contribution of the nonlinear term in (46) up to leading term involving $\kappa$. Obviously, the terms involving $\kappa^2$ and $\kappa$ with higher exponents can be easily incorporated into (50) using the expansion (49).

**Remark 3.** There are some special cases when the integral (with correctly-specified parameters) arising in (48) is expressed in elementary functions. One may discuss whether such cases are interesting from the applicability point of view.

**Example 1.** Let us consider a layer of PEM at the steady-state with surfaces at $x = x_0 = 0$ and $x = L > 0$, both permeable. In particularly, it might be a human tissue (interstitium) with slightly negative and constant interstitial (hydrostatic) pressure of -1 mmHg, containing two solutes of different molecular size and fluid space available for their transport, for example, glucose that is transported within $\theta_1$ and $\theta_2$, and albumin, which transport is restricted to $\theta_2$ and $\theta_2 = 0.4\theta_F$, i.e. $\alpha = 0.4$ [30]. Let us assume that at this initial steady-state the concentration of glucose and albumin is 6 and 0.4 mmoL/L, respectively with $\sigma_1 = 0.0035$ and $\sigma_2 = 0$.

In this example, we are interested in the changes of the tissue size (deformations) triggered by the immerse of external surface of PEM (surface at $x = L$) into some fluid resulting in the change of effective pressure $P^*$ exerted at this surface (caused by changes of hydrostatic pressure and/or solutes concentration at external side of the PEM layer). It would lead to the possible deformations of the tissue, displacement of the external surface from $x = L$ to $x = L_S$ at the new steady-state (change of tissue width from $L$ to $L_S$). Let us consider hypertonic fluid (used in medical treatment called peritoneal dialysis) containing glucose 3.86% and no albumin. Namely, we assume that fluid exert hydrostatic pressure of 40 mmHg, contains glucose at concentration of 170 mmoL/L, and no albumin. Let us consider the width of deformated tissue be 1 cm, and denote by $X$ initial position in PEM, and by $x$ at the new steady-state (after deformation). Therefore, $U(x) = x - X$.

For calculations, we applied the formula (50) with the function $G(x)$ given in (46). In order to avoid cumbersome expressions, we set $\gamma_0 + \gamma_1 = \sigma_1$ and $\gamma_2 = \alpha\sigma_2$, therefore a cubic polynomial is obtained for the tissue displacement $U(x)$ as a function of the present position $x$. The expression of the present state position as a function of initial position was caluclated using definition of the displacement i.e. $U(x) = x - X$ We note that it would be a transcendent equation without the above restrictions on the parameters $\gamma_i$, $i = 0, 1, 2$.

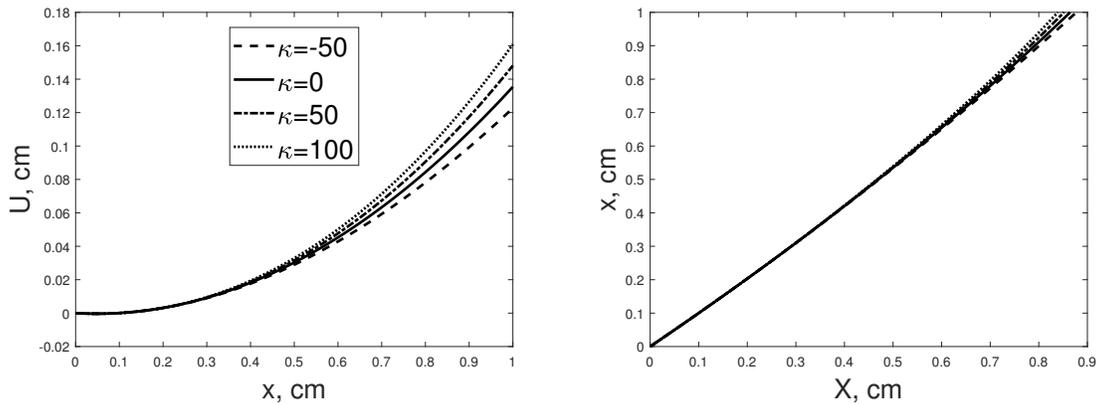

Figure 2: The profiles of the healthy tissue ($\lambda^* = 100$) displacement $U(x)$ as a function of the present position $x$ i.e. when new steady-state is obtained (left panel) and the corresponding profiles of the present state position in PEM $x = U + X$ as a function of initial state position $X$ (right panel) for for the nonlinear stress tensor with parameter $\kappa = -50, 0, 50, 100$.

In Figure 2, left panel we present the profiles of tissue displacement $u$ as a function of present position $x$ for different values of $\kappa$ assuming elastic modulus parameter $\lambda^* = 100$, which corresponds to the healthy tissue. To better visualize the observed changes within the tissue in Figure 2, right panel, we present the corresponding current position within the PEM as a function of initial position. In case of linear stress tensor i.e. $\kappa = 0$ the would change its size by 0.135 cm, this in case of nonlinear stress tensor of quadratic form with $\kappa = 100$, the observed displacement would be 0.161, and 0.123 in case of $\kappa = -50$, see Figure 2.

In case of more stiffness material (such as tumour tissue), the corresponding impact of the nonlinear term is less pronounced. In Figure 3, we present results obtained assuming that elastic modulus $\lambda^* = 700$, and $\kappa = -50, 0, 50,$ and $100$. One notes that all the curves practically coincide independently of the parameter $\kappa$.

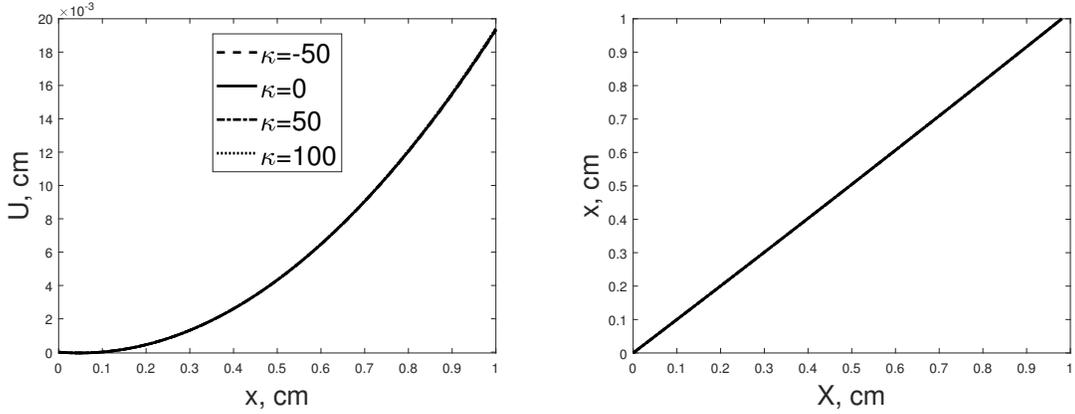

Figure 3: The profiles of the tumour tissue ($\lambda^* = 700$) displacement $U(x)$ as a function of the present position $x$ i.e. when new steady-state is obtained (left panel) and corresponding profiles of the present position $x = U + X$ as a function of initial state position $X$ (right panel) for the nonlinear stress tensor with parameter $\kappa = -50, 0, 50, 100$.

## 4  Lie symmetries and time-dependent exact solutions

Lie symmetry of the PDE system (33)–(38) depends essentially on values of the parameters arising therein. From the very beginning we restrict ourselves on the values of the parameters which do not affect the biological meaning of the system and do not lead to such simplifications, which have been studied elsewhere. So, we assume that several parameters are non-zero:

$$k\alpha\rho_F^0 D_i S_i \neq 0, \ i = 1, 2. \tag{51}$$

**Theorem 1** *The principal algebra of invariance of system (33)–(38) is a infinitely-dimensional Lie algebra generated by the Lie symmetry operators*

$$X_1 = \partial_t, \ X_2 = \partial_x, \ X_3 = \partial_u, \ X_4 = g(t)\partial_{p^*}, \tag{52}$$

*where $g(t)$ is an arbitrary function. Hereafter the notations $\partial_z \equiv \frac{\partial}{\partial z}, \ (z = t, x, u, ...)$ are used.*

We remind the reader that the notion of the principal algebra of invariance (see, e.g., [35] ) means that a given PDE (system of PDEs) admits this algebra for arbitrary set of parameters arising in the PDE (system of PDEs) in question. Moreover, this algebra of invariance cannot be extended by any other Lie symmetry without a restriction on parameters of the given PDE (system of PDEs).

It turns out that the values of the parameters $\gamma_0$, $\gamma_1$, $\gamma_2$ and $\kappa$ play essential role when one looks for further extensions of the principal algebra (52).

**Theorem 2** *The principal algebra (52) of the PDE system (33)–(38) with the coefficient restriction (51) admits the extensions listed in Table 1. There are not other Lie symmetries, which the system admits under restriction (51).*

**Proof** of Theorem 1 and Theorem 2 is based on the infinitesimal criteria of invariance, which was formulated by S. Lie in his pioneering works [31, 32]. In the case of a system of PDEs of arbitrary order, this criteria can be found, e.g., in [33] (see Section 1.2.5).

Because system (33) consists of 6 PDEs of the second order, we start from the linear first-order operator of the form

$$X = \xi_0 \partial_t + \xi_1 \partial_x + \eta_1 \partial_u + \eta_2 \partial_\rho + \eta_3 \partial_{p^*} + \eta_4 \partial_{\theta_F} + \eta_5 \partial_{c_1} + \eta_6 \partial_{c_2}. \tag{53}$$

Here the operator coefficients $\xi_0$, $\xi_1$ and $\eta_i$, $i=1,\ldots,6$ are arbitrary (at the moment) smooth functions of independent and dependent variables.

Operator (53) is a Lie symmetry (the equivalent terminology is 'operator of Lie's invariance' and 'point symmetry') of system (33) provided the following equalities take place

$$\begin{aligned}
&X^{(2)}\Big(2u_{tx} - kp^*_{xx}\Big) = 0, \\
&X^{(2)}\Big(\rho_t + \rho_x u_t - k(\rho_F^0 - \rho)p^*_{xx}\Big) = 0, \\
&X^{(2)}\Big(\theta_{Ft} + \theta_{Fx} u_t - k\left(1 - \theta_F\right)p^*_{xx}\Big) = 0, \\
&X^{(2)}\Big((\theta_F c_1)_t + (\theta_F c_1)_x u_t + 2\theta_F c_1 u_{tx} - D_1 c_{1,xx} - kS_1(c_1 p^*_x)_x\Big) = 0, \\
&X^{(2)}\Big((\theta_F c_2)_t + (\theta_F c_2)_x u_t + 2\theta_F c_2 u_{tx} - D_2 c_{2,xx} - kS_2(c_2 p^*_x)_x\Big) = 0, \\
&X^{(2)}\Big(\rho u_{tt} + \rho_t u_t + \rho u_t u_{tx} - \lambda^* u_{xx} - 2\kappa u_x u_{xx} + \\
&\qquad p^*_x - (\gamma_0 + \gamma_1 - \sigma_1)RTc_{1,x} - (\gamma_2 - \alpha\sigma_2)RTc_{2,x}\Big) = 0
\end{aligned} \tag{54}$$

for each solution $\Big(u(t,x),\ \rho(t,x),\ p^*(t,x),\ \theta_F(t,x),\ c_1(t,x),\ c_2(t,x)\Big)$ of the PDE system (33)–(38). Here $X^{(2)}$ is the second-order prolongation of the operator $X$, which is again a linear first-order operator with coefficients defined by the well-known formulae via the first- and second-order derivatives of unknown coefficients $\xi_0$, $\xi_1$ and $\eta_i$ (see, e.g., [33], Section 1.2.1).

Substituting the expression for $X^{(2)}$ into (54) and having done long calculations (see typical examples in the books [33–35]), one arrives at a linear system of PDEs, which is called system of determining equations(DEs), to find the functions $\xi_0$, $\xi_1$ and $\eta_i$. Because the system of determining equations is always linear and overdetermined (the number of PDEs is larger than that of unknown functions), usually its general solution can be constructed by straightforward calculations. Moreover one may use the computer algebra packages such as Maple and Mathematica for such purposes. Finally, substituting the functions $\xi_0$, $\xi_1$ and $\eta_i$ into (53), one easily constructs the set of Lie symmetry operators producing the Lie algebra of invariance. In the case of (54), the system of DEs has the general solution

$$\xi_0 = \xi_0^0, \ \xi_1 = \xi_1^0, \ \eta_1 = \eta_1^0, \ \eta_2 = 0, \ \eta_3 = g(t), \ \eta_4 = \eta_5 = \eta_6 = 0, \tag{55}$$

where $\xi_0^0$, $\xi_1^0$ and $\eta_1^0$ are arbitrary constants, $g(t)$ is an arbitrary smooth function. So, operator (53) with coefficients (55) produces four linearly independent operators (52) generating the principal algebra of invariance of system (33)–(38).

However, the system in question contains several parameters and some of them can vanish. As a result, the general solution of the system of DEs may be different for some correctly-specified parameters, hence such special cases must be identified. In the case of system (33)–(38), it was shown that the condition

$$\kappa(\gamma_0 + \gamma_1 - \sigma_1)(\gamma_2 - \alpha\sigma_2) = 0$$

leads to extensions of the Lie algebra (52). Thus, six different restrictions on the coefficients arising in the above condition were analyzed. The result is presented in Table 2.

The sketch of the proof is completed.

**Remark 4.** Theorem 2 can be easily reformulated for the initial system (26)–(31) because both system are related via transformation (32). As a result, Table 2 will be transformed in a quite similar that with the only differences that three Lie symmetries take the following forms

$$g(t)\partial_{p^*} \to g(t)\partial_p, \ c_1\partial_{c_1} \to c_1\Big(\partial_{c_1} + T_1\partial_p\Big), \ c_2\partial_{c_2} \to c_2\Big(\partial_{c_2} + \alpha T_2\partial_p\Big).$$

Table 2: Lie symmetries of system (33)–(38)

| | Restrictions | Lie symmetries |
|---|---|---|
| 1. | $\kappa \neq 0,\ \gamma_0 + \gamma_1 = \sigma_1$ | $X_1 = \partial_t,\ X_2 = \partial_x,\ X_3 = \partial_u,\ X_4 = g(t)\partial_{p^*},$ $X_5 = c_1 \partial_{c_1}$ |
| 2. | $\kappa \neq 0,\ \gamma_2 = \alpha \sigma_2$ | $X_1 = \partial_t,\ X_2 = \partial_x,\ X_3 = \partial_u,\ X_4 = g(t)\partial_{p^*},$ $X_6 = c_2 \partial_{c_2}$ |
| 3. | $\kappa \neq 0,\ \gamma_0 + \gamma_1 = \sigma_1$ $\gamma_2 = \alpha \sigma_2$ | $X_1 = \partial_t,\ X_2 = \partial_x,\ X_3 = \partial_u,\ X_4 = g(t)\partial_{p^*},$ $X_5 = c_1 \partial_{c_1},\ X_6 = c_2 \partial_{c_2}$ |
| 4. | $\kappa = 0,,\ \gamma_0 + \gamma_1 = \sigma_1$ | $X_1 = \partial_t,\ X_2 = \partial_x,\ X_3 = \partial_u,\ X_4 = g(t)\partial_{p^*},$ $X_5 = c_1 \partial_{c_1},\ X_7 = x \partial_u$ |
| 5. | $\kappa = 0,\ \gamma_2 = \alpha \sigma_2$ | $X_1 = \partial_t,\ X_2 = \partial_x,\ X_3 = \partial_u,\ X_4 = g(t)\partial_{p^*},$ $X_6 = c_2 \partial_{c_2},\ X_7 = x \partial_u$ |
| 6. | $\kappa = 0,\ \gamma_0 + \gamma_1 = \sigma_1$ $\gamma_2 = \alpha \sigma_2$ | $X_1 = \partial_t,\ X_2 = \partial_x,\ X_3 = \partial_u,\ X_4 = g(t)\partial_{p^*},$ $X_5 = c_1 \partial_{c_1},\ X_6 = c_2 \partial_{c_2},\ X_7 = x \partial_u$ |
| | | |

# 5 Time-dependent solutions of the model

## 5.1 Reduction the model to systems of ODEs

According to the classification presented in the previous section, the widest Lie symmetry of the PDE system (33)–(38) occurs in Case 6 (see Table 2). So, equation (38) reduces to the form

$$\rho u_{tt} + \rho_t u_t + \rho u_t u_{tx} = \lambda^* u_{xx} - p_x^*. \tag{56}$$

So, the Lie algebra of invariance of system (33)–(37) and (56) is generated by the operators

$$\begin{aligned} &X_1 = \partial_t,\ X_2 = \partial_x,\ X_3 = \partial_u,\ X_4 = g(t)\partial_{p^*},\\ &X_5 = c_1 \partial_{c_1},\ X_6 = c_2 \partial_{c_2},\ X_7 = x\partial_u. \end{aligned} \tag{57}$$

Although it is an infinitely-dimensional Lie algebra, its structure is rather simple because there are only two non-zero Lie brackets commutators, $[X_1, X_4] = g'(t)\partial_{p^*}$ i.e. $X_4$ with $g(t) \to g'(t)$, and $[X_2, X_7] = X_3$. All other commutators give zeros. From the algebraic point of view, the Lie algebra (57) is a semidirect sum of two Abelian algebras with the basic operators $X_2, \ldots, X_6$ and $X_1, X_7$, respectively.

Using Lie algebra of invariance (56), one has several possibilities to reduce system (33) to systems of ODEs in order to find exact solutions with different structures. For example, using the Lie symmetry $\partial_t$ one immediately identifies ansatz (40) leading to steady-state solutions. However, Lie symmetries allow us to construct ansätze with more complicated structures. It is well-known that each ansatz reduces system (33)–(37) and (56) to a system of ODEs.

In this section, we are looking for time-dependent solutions of system (33)–(37) and (56). First of all, we remind the reader that any linear combination of Lie symmetries from (57) is also a Lie symmetry of the system in question. The most general linear combination involves all the operators and reads as

$$Y = \alpha_0 \partial_t + \alpha_1 \partial_x + \alpha_2 \partial_u + \alpha_4 g(t)\partial_{p^*} + \alpha_5 c_1 \partial_{c_1} + \alpha_6 c_2 \partial_{c_2} + \alpha_3 x\partial_u \tag{58}$$

where $\alpha_i$, $i = 0\ldots 7$ are arbitrary parameters.

The operator $Y$ (58) contains seven arbitrary parameters $\alpha_i$, $i = 0\ldots 6$ and the function $g(t)$. A common problem arises how to find such linear combinations of these operators, which lead to inequivalent reductions of system in question and therefore to essentially different exact solutions. In fact, depending on values of these parameters, different ansätze can be derived. In order to identify essentially different ansätze, there are two approaches (see [35], Section 1.3). The first one is to construct all inequivalent solutions of the corresponding system of characteristic equations, the second one is to solve a pure algebraic problem about finding non-conjugated subalgebras of Lie algebra (57). Here we use the first approach, which does not involve a sophisticated technique and one is based on direct analysis of a system of characteristic equations (the terminology 'invariance surface condition' is also used). So, the system of characteristic equations corresponding to (58) is

$$\frac{dt}{\alpha_0} = \frac{dx}{\alpha_1} = \frac{du}{\alpha_2 + \alpha_3 x} = \frac{dp^*}{\alpha_4 g(t)} = \frac{dc_1}{\alpha_5 c_1} = \frac{dc_2}{\alpha_6 c_2}. \tag{59}$$

Now one realizes that the first equation $\frac{dt}{\alpha_0} = \frac{dx}{\alpha_1}$, depending on $\alpha_0$ and $\alpha_1$ leads only to two essentially different form of the invariant variable $\omega(t,x)$, namely *Case (i)* $\omega = t$, if $\alpha_0 = 0$, $\alpha_1 \neq 0$ and *Case (ii)* $\omega = x - vt$, $v = \frac{\alpha_1}{\alpha_0}$ if $\alpha_0 \neq 0$. The case $\alpha_0 = \alpha_1 = 0$ is excluded in what follows because one does not lead to the invariant variable $\omega$ as a function of $t$ and $x$ but to an overdetermined system of PDEs for finding the functions $u$, $\rho$, $p^*$, $\theta_F$, $c_1$ and $c_2$.

**Case (i).** Taking into account that $\alpha_1 \neq 0$, system (59) can be rewritten as follows

$$(u_1 + 2u_2 x)dx = du, \quad \phi_3(t)dx = dp^*, \quad -w_1 dx = \frac{dc_1}{c_1}, \quad -w_2 dx = \frac{dc_2}{c_2}, \tag{60}$$

where $u_1 = \frac{\alpha_2}{\alpha_1}$, $2u_2 = \frac{\alpha_3}{\alpha_1}$, $\phi_3(t) = \frac{\alpha_4}{\alpha_1}g(t)$, $w_1 = -\frac{\alpha_5}{\alpha_1}$, and $w_2 = -\frac{\alpha_6}{\alpha_1}$.

Integrating each equation from (60) and taking into account the dependence of arbitrary constants depend on the invariant variable $\omega = t$, one arrives at the ansatz

$$u = u_1 x + u_2 x^2 + \phi_1(t), \quad \rho = \phi_2(t), \quad p^* = p_0(t) + \phi_3(t)x,$$
$$\theta_F = \phi_4(t), \quad c_1 = e^{-w_1 x}\phi_5(t), \quad c_2 = e^{-w_2 x}\phi_6(t), \tag{61}$$

where $\phi_i(t)$, $i = 1, \ldots, 6$ are to-be-determined functions.

Substituting ansatz (61) into system (33)–(37) and (56) and making straightforward calculations, we arrive at the following ODE system for finding $\phi_i$, $i = 1, \ldots, 6$:

$$\dot{\phi}_2 = 0, \quad \dot{\phi}_4 = 0,$$
$$\dot{\phi}_5 - w_1\dot{\phi}_1\phi_5 = w_1^2 D_1 \phi_5 - w_1 k S_1 \phi_3 \phi_5,$$
$$\dot{\phi}_6 - w_2\dot{\phi}_1\phi_6 = w_2^2 D_2 \phi_6 - w_2 k S_2 \phi_3 \phi_6, \tag{62}$$
$$\phi_2 \ddot{\phi}_1 = 2u_2 - \phi_3.$$

(here the dot means the time derivative). Notably equation (33) vanishes if one applies ansatz (61).

**Case (ii).** A similar analysis as in Case (i) shows that the form of the corresponding ansatz depends essentially on the parameter $v$. In fact, having $v = \alpha_1 = 0$, we obtain $\omega = x$, therefore system (59) can be rewritten as follows

$$(u_0 + u_1 x)dt = du, \ \dot{p}_0(t)dt = dp^*, \ -v_1 dt = \frac{dc_1}{c_1}, \ -v_2 dt = \frac{dc_2}{c_2}, \tag{63}$$

where $u_0 = \frac{\alpha_2}{\alpha_0}$, $u_1 = \frac{\alpha_3}{\alpha_0}$, $\dot{p}_0(t) = \frac{\alpha_4}{\alpha_0} g(t)$, $v_1 = -\frac{\alpha_5}{\alpha_0}$, and $v_2 = -\frac{\alpha_6}{\alpha_0}$. Integrating the ODE system (63) and taking into account that arbitrary constants depend on the invariant variable $\omega = x$, one arrives at the ansatz

$$u = (u_0 + u_1 x)t + \phi_1(x), \ \rho = \phi_2(x), \ p^* = p_0(t) + \phi_3(x),$$
$$\theta_F = \phi_4(x), \ c_1 = e^{-v_1 t}\phi_5(x), \ c_2 = e^{-v_2 t}\phi_6(x), \tag{64}$$

where $\phi_i(x)$, $i = 1, \ldots, 6$ are to-be-determined functions.

**Remark 5.** In the case $u_0 = u_1 = v_1 = v_2 = p_0(t) = 0$, this ansatz reduces to that used in Section 3 for finding steady-state solutions.

Assuming $v \neq 0$, we obtain $\omega = x - vt$. In this case, the system of characteristic equations (59) produces the ansatz

$$u = u_1 x + u_2 x^2 + \phi_1(\omega), \ \rho = \phi_2(\omega), \ p^* = p_0(t) + \phi_3(\omega),$$
$$\theta_F = \phi_4(\omega), \ c_1 = e^{-v_1 t}\phi_5(\omega), \ c_2 = e^{-v_2 t}\phi_6(\omega), \tag{65}$$

with unknown functions $\phi_i(\omega)$, $i = 1, \ldots, 6$.

**Remark 6.** In the case $u_i = v_i = p_0(t) = 0$ $(i = 1, 2)$, this ansatz reduces to the well-known substitution for finding plane wave solutions (in particular, traveling fronts).

By substitution of ansätze (64) and (65) into system (33)–(37) and (56), we reduce the given PDE system to ODE systems. As a result, we obtain the system

$$\begin{aligned}
& k\phi_3'' = 2u_1, \\
& (u_0 + u_1 x)\phi_2' = 2u_1\left(\rho_F^0 - \phi_2\right), \\
& (u_0 + u_1 x)\phi_4' = 2u_1(1 - \phi_4), \\
& (2u_1 - v_1)\phi_4\phi_5 + (u_0 + u_1 x)(\phi_4\phi_5)' = v_1^2 D_1 \phi_5'' + kS_1(\phi_5\phi_3')', \\
& (2u_1 - v_1)\phi_4\phi_6 + (u_0 + u_1 x)(\phi_4\phi_6)' = v_2^2 D_2 \phi_6'' + kS_2(\phi_6\phi_3')', \\
& u_1(u_0 + u_1 x)\phi_2 = \lambda^* \phi_1'' - \phi_3',
\end{aligned} \tag{66}$$

if $v = 0$ and the system

$$
\begin{aligned}
&-2v\phi_1'' = k\phi_3'', \\
&(1 + \phi_1')\phi_2' = -\tfrac{k}{v}\left(\rho_F^0 - \phi_2\right)\phi_3'' \\
&(1 + \phi_1')\phi_4' = -\tfrac{k}{v}\left(1 - \phi_4\right)\phi_3'', \\
&v_1\phi_4\phi_5 + v(1 + \phi_1')(\phi_4\phi_5)' + 2v\phi_4\phi_5\phi_1'' = -\left(D_1\phi_5'' + kS_1\left(\phi_5\phi_3'\right)'\right), \\
&v_2\phi_4\phi_6 + v(1 + \phi_1')(\phi_4\phi_6)' + 2v\phi_4\phi_6\phi_1'' = -\left(D_2\phi_6'' + kS_2\left(\phi_6\phi_3'\right)'\right), \\
&v^2(\phi_2\phi_1')' + v^2\phi_2\phi_1'\phi_1'' = \lambda^*\phi_1'' - \phi_3',
\end{aligned}
\tag{67}
$$

if $v \neq 0$ (hereinafter primes denote the differentiation w.r.t. variable $\omega$).

## 5.2 Construction of exact solutions

This subsection is devoted to solving the reduced systems (62), (66) and (67). Having exact solutions of these ODE systems, one easily obtains solutions of the model, i.e. the nonlinear PDE system (33)–(37) and (56).

Let us start from the ODE system (62). Because we have 5 equations for 6 unknown functions, the function $\phi_1(t)$ can be thought as an arbitrary smooth. So, solving the first two equations, one immediately obtains $\phi_2(t) = \rho^0$ and $\phi_4(t) = \theta_F^0$ (hereinafter $\rho^0$ and $\theta_F^0$ are arbitrary constant, which should be positive if one looks for exact solutions with a physical/biological/chemical interpretation). Now we find the unction $\phi_3(t)$ from the last equation of system (62) and substitute one into the 3rd and 4th equations. As a result, we arrive at two linear ODEs for $\phi_5(t)$ and $\phi_6(t)$, which can be readily solved. Finally, using ansatz (61), we obtain the following multiparameter family of exact solutions of the nonlinear model (33)–(37) and (56):

$$
\begin{aligned}
u(t, x) &= u_1 x + u_2 x^2 + f(t), \\
\rho(t, x) &= \rho^0, \ \theta_F(t, x) = \theta_F^0, \\
p^*(t, x) &= (2u_2 - \rho^0 \ddot{f}(t))x + p_0(t), \\
c_1(t, x) &= A_1 \exp\left(w_1\left(-x + v_1 t + f(t) + \rho^0 k S_1 \dot{f}(t)\right)\right), \\
c_2(t, x) &= A_2 \exp\left(w_2\left(-x + v_2 t + f(t) + \rho^0 k S_2 \dot{f}(t)\right)\right),
\end{aligned}
\tag{68}
$$

where $u_i,\ w_i,\ A_i, \rho^0$ and $\theta_F^0\ (i = 1, 2)$ are arbitrary constants, $p_0(t)$ and $f(t)$ are arbitrary smooth functions, while $v_i = w_i D_i - 2k u_2 S_i$.

Although the ODE system (66) possesses much more complicated structure than (62), one still can be fully integrated, however, its solutions are expressed via special functions in the general case. It can be easily noted that the process of integration of system (66) essentially depends on the parameter $u_1$, therefore one should separately examine *Case (i)* $u_1 \neq 0$ and *Case (ii)* $u_1 = 0$.

**Case (i)**. In this case, integration of the first three equations from (66) leads to

$$\begin{aligned}
\phi_3 &= \tfrac{u_1}{k} x^2 + p_1 x + p_0, \\
\phi_2 &= \rho_F^0 + \rho^1 (x + x_0)^{-2}, \\
\phi_4 &= 1 - \theta^1 (x + x_0)^{-2},
\end{aligned} \qquad (69)$$

where $p_1$, $p_0$, $\rho^1$ and $\theta^1$ are arbitrary constants (the latter should be positive for the applicability point of view), while $x_0 = \tfrac{u_0}{u_1}$. Having $\phi_2$ and $\phi_3$, the function $\phi_1$ is determined from the last equation of system (66):

$$\phi_1 = \tfrac{u_1^2 \rho^1}{\lambda^*}(x+x_0)\ln|x+x_0| + \tfrac{u_1}{6k\lambda^*}(2+u_1 k\rho_F^0)(x+x_0)^3 + \tfrac{1}{2k\lambda^*}(kp_1 - 2u_0)(x+x_0)^2 + U_1 x + U_0, \quad (70)$$

where $U_1$ and $U_0$ are two additional constants, which can be skipped without losing a generality (see subsection 5.1).

Finally, we need to solve two remaining equations in order to find the functions $\phi_5$ and $\phi_6$. Using the functions $\phi_3$ and $\phi_4$ from (69), these equations reduce to the form

$$\begin{aligned}
D_1 \phi_5'' + \Big(u_1 \theta^1 (x+x_0)^{-1} + u_1(2S_1 - 1)x + \beta_{11}\Big)\phi_5' + \Big(-v_1 \theta^1 (x+x_0)^{-2} + \beta_{12}\Big)\phi_5 &= 0, \\
D_2 \phi_6'' + \Big(u_1 \theta^1 (x+x_0)^{-1} + u_1(2S_2 - 1)x + \beta_{21}\Big)\phi_6' + \Big(-v_2 \theta^1 (x+x_0)^{-2} + \beta_{22}\Big)\phi_6 &= 0,
\end{aligned} \qquad (71)$$

where $\beta_{i1} = kp_1 S_i - u_0$ and $\beta_{i2} = 2u_1(S_i - 1) + v_i$ ($i = 1, 2$).

Each equation in (71) is a linear second-order ODE and has the general solution in the form of a linear combination of two hypergeometric functions [43]. Thus, using ansatz (64), we obtain the following multiparameter family of exact solutions of the nonlinear model (33)–(37) and (56):

$$\begin{aligned}
u(t,x) &= u_1(x + x_0)t + \tfrac{u_1^2 \rho^1}{\lambda^*}(x+x_0)\ln|x+x_0| + \tfrac{u_1}{6k\lambda^*}(2+u_1 k\rho_F^0)(x+x_0)^3 \\
&\quad + \tfrac{1}{2k\lambda^*}(kp_1 - 2u_0)(x+x_0)^2 + U_1 x + U_0 \\
\rho(t,x) &= \rho_F^0 + \rho^1(x+x_0)^{-2}, \quad \theta_F(t,x) = 1 - \theta^1(x+x_0)^{-2}, \\
p^*(t,x) &= \tfrac{u_1}{k} x^2 + p_1 x + p_0(t), \\
c_1(t,x) &= e^{-v_1 t}\Big(A_{11} F_{11}(x) + A_{12} F_{12}(x)\Big), \\
c_2(t,x) &= e^{-v_1 t}\Big(A_{21} F_{21}(x) + A_{22} F_{22}(x)\Big),
\end{aligned} \qquad (72)$$

where $F_{ij}(x)$ are correctly-specified (by the coefficients in (71)) hypergeometric functions and $A_{ij}$ are arbitrary constants (other parameters are described above).

In the general case, hypergeometric function are not useful for practical applications, therefore it is important to look for particular cases leading to simpler functions. Let us present an example. Consider the first equation in (71) with $S_1 = 1/2$ and $\beta_{11} = 0$ (i.e. $u_0 = kp_1/2$). It is a realistic choice because $u_0$ and $u_1$ are arbitrary, while $S_1 < 1$ is a typical restriction on the sieving coefficient (see Table 1). Under above restrictions, this equation takes the form of the generalized Bessel equation

$$(x+x_0)^2 \phi_5'' + \chi(x+x_0)\phi_5' - \left(\frac{v_1\theta^1}{D_1} + \frac{u_1-v_1}{D_1}(x+x_0)^2\right)\phi_5 = 0, \quad \chi = \frac{u_1\theta^1}{D_1}. \tag{73}$$

Thus, depending on the value of $\frac{u_1-v_1}{D_1}$ one obtains the general solution

$$\phi_5 = \begin{cases} A_1(x+x_0) + A_2(x+x_0)^{-\chi}, \ u_1 = v_1 \\ A_1 J_\nu\Big(|B|(x+x_0)\Big) + A_2 Y_\nu\Big(|B|(x+x_0)\Big), \ B^2 = \frac{v_1-u_1}{D_1} > 0, \\ A_1 I_\nu\Big(|B|(x+x_0)\Big) + A_2 K_\nu\Big(|B|(x+x_0)\Big), \ B^2 = \frac{u_1-v_1}{D_1} > 0, \end{cases} \tag{74}$$

where $A_1$ and $A_2$ are arbitrary constants and $\nu = \frac{1}{2}\sqrt{(\chi-1)^2 + \frac{4v_1\theta^1}{D_1}}$. The notations $J_\nu$ and $Y_\nu$ stand for the Bessel functions of the first and third kind, respectively, while those $I_\nu$ and $K_\nu$ stand for the modified Bessel functions of the first and third kind, respectively (see, e.g. [44]).

**Case (ii)**. In this case, the integration process of system (66) is rather trivial. As a result, one obtains the multiparameter family of exact solutions of the nonlinear model (33)–(37) and (56)

$$\begin{aligned}
u(t,x) &= u_0 t + \tfrac{p_1}{2\lambda^*}x^2 + U_1 x + U_0 \\
\rho(t,x) &= \rho^0, \ \theta_F(t,x) = \theta_F^0, \\
p^*(t,x) &= p_1 x + p_0(t), \\
c_1(t,x) &= e^{-v_1 t}\Big(A_{11}f_{11}(x) + A_{12}f_{12}(x)\Big), \\
c_2(t,x) &= e^{-v_1 t}\Big(A_{21}f_{21}(x) + A_{22}f_{22}(x)\Big),
\end{aligned} \tag{75}$$

where $f_{ij}(x)$ are the fundamental systems of solutions of the linear ODEs with constant coefficients

$$\begin{aligned}
D_1\phi_5'' + (kp_1 S_1 - u_0\theta_F^0)\phi_5' + v_1\theta_F^0\phi_5 &= 0, \\
D_2\phi_6'' + (kp_1 S_2 - u_0\theta_F^0)\phi_6' + v_2\theta_F^0\phi_6 &= 0.
\end{aligned} \tag{76}$$

Obviously, ODEs (76) produce essentially different solutions depending on coefficients. In particular, periodic and quasiperiodic functions can be derived.

In conclusion of this section, we present a brief analysis of the ODE system (67). In contrast to the ODE systems (62) and (66), we were unable to solve (67) in general case. However, it can be noted the special case $\phi_1' = -1$ when the system essentially simplifies. In this case, one readily identifies that $\phi_1 = U_0 - \omega$, $\phi_2 = \rho_0 + \frac{p_1}{v^2}\omega$, $\phi_3 = p_1\omega + p_0$, while $\phi_4$ is an arbitrary smooth function because the third equation in (67) vanishes. As a result, the remaining equations in system (67) take the form

$$D_1\phi_5'' + kp_1 S_1 \phi_5' + v_1\phi_4\phi_5 = 0,$$
$$D_2\phi_6'' + kp_1 S_2 \phi_6' + v_2\phi_4\phi_6 = 0. \tag{77}$$

Each equation in (77) is a linear second-order ODE with the variable coefficient $\phi_4(\omega)$ and its general solution depends on its form. In particular, system (77) coincides with (76) if $\phi_4(\omega) = \theta_F^0$ and $u_0 = 0$.

Thus, the following multiparameter family of exact solutions of the nonlinear model (33)–(37) and (56) is derived

$$u(t,x) = u_2 x^2 + u_1 x + u_0 + vt$$
$$\rho(t,x) = \rho_0 + \frac{p_1}{v^2}(x - vt),,$$
$$p^*(t,x) = p_1 x + p_0(t), \tag{78}$$
$$c_1(t,x) = e^{-v_1 t}\Big(A_{11} f_{11}(x) + A_{12} f_{12}(x)\Big),$$
$$c_2(t,x) = e^{-v_1 t}\Big(A_{21} f_{21}(x) + A_{22} f_{22}(x)\Big),$$

where $f_{ij}(x)$ are the fundamental systems of solutions of the linear ODEs (77). The function $\theta_F$ is an arbitrary smooth function of $\omega = x - vt$.

**Example 2.** Now we consider the layer of PEM of the width $L$ like in Example 1. However, in contrast to Example 1, we allow its evolution in time, assuming that the layer is shrinking because an external force (load) is acted on the external surface $x = L$. Our aim is to demonstrate applicability of the exact solutions obtained for describing the layer deformation and solute transport, assuming that there is the solute consisting of molecules of the equal size, i.e. $c_1 = c(t,x)$, $c_2 = 0$. We do not go into the detailed description of these processes (this lies beyond scopes of this study) but it turns out that a relatively simple solution describes those at least qualitatively.

It is naturally to assume that the initial displacement is zero, i.e. $u(0,x) = 0$, $x \in [0, L]$, and the left end of the layer is fixed, i.e. $u(t,0) = 0$. A simple analysis of (72) shows that this occurs when

$$\rho^1 = U_1 = U_0 = 0, \ u_1 = -\frac{2}{k\rho_F^0}, \ u_0 = \frac{kp_1}{2}. \tag{79}$$

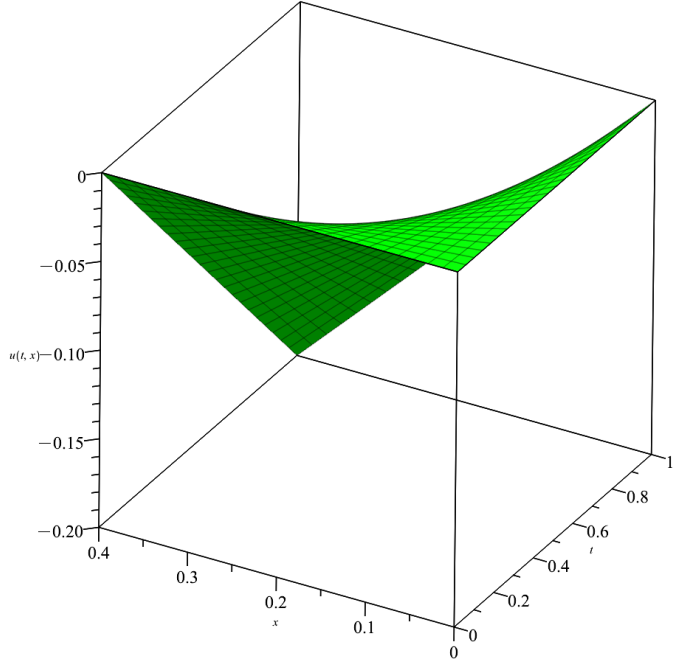

Figure 4: The function $u(t,x)$ represents displacement (deformation) of the layer of PEM. This 3D plot and those presented in Fig. 5 and 6 are drawing using formulae (82). The parameters arising in (82) were taken as the following non-dimensional values $k = 2$; $\rho_F^0 = 2$; $RT = 0.2$; $P_0 = 1$; $p_1 = -1$; $A_1 = 1$; $x_0 = 2$; $L = 0.4$; $\theta^1 = 4$; $D_1 = 1.5$; $\chi = -\frac{2\theta^1}{k\rho_F^0 D_1} = -\frac{4}{3}$.

Two widely used boundary conditions for pressure and concentrations are

$$x = 0: \quad p^* = P_0, \quad \frac{\partial c}{\partial x} = 0. \tag{80}$$

The first condition in (80) immediately leads to $p_0(t) = P_0$ (see (72)). To apply the second boundary condition we need to specify the formula for $c(t,x)$. Setting additional $S_1 = 1/2$ and $u_1 = v_1$, one realizes that the first formula from (74) can be used. Thus, the concentration is

$$c(t,x) = e^{-u_1 t}\Big(A_1(x+x_0) + A_2(x+x_0)^{-\chi}\Big), \quad \chi = \frac{u_1 \theta^1}{D_1}. \tag{81}$$

The no-flux condition is satisfied provided $A_2 = \frac{A_1}{\chi} x_0^{1+\chi}$.

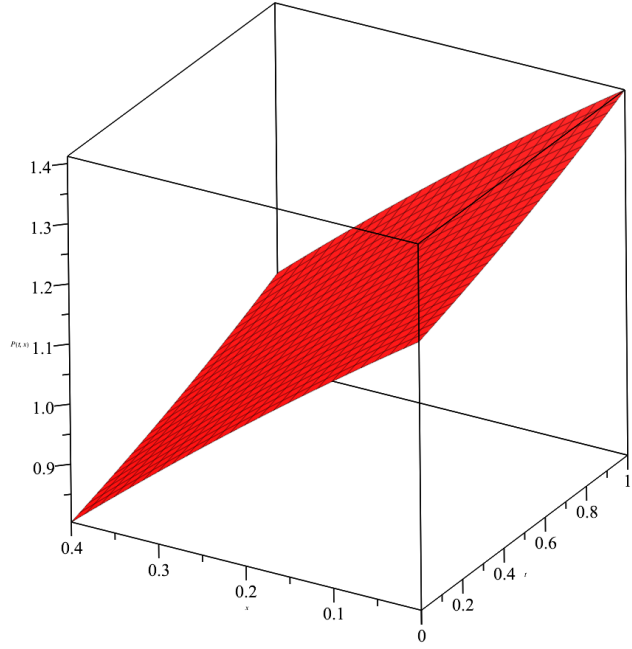

Figure 5: The function $p(t,x)$ represents the hydrostatic pressure in the PEM layer.

Thus, summarizing the above results, taking into account formula (32) and using the exact solution (72), we arrive at the solution

$$u(t,x) = -\frac{2}{k\rho_F^0}tx, \quad \rho(t,x) = \rho_F^0, \quad \theta_F(t,x) = 1 - \frac{\theta^1}{(x+x_0)^2},$$
$$p(t,x) = P_0 + p_1 x - \frac{2}{k^2\rho_F^0}x^2 + \frac{1}{2}RTA_1\exp(\frac{2}{k\rho_F^0}t)\Big((x+x_0) + \frac{1}{\chi}x_0^{1+\chi}(x+x_0)^{-\chi}\Big), \quad (82)$$
$$c(t,x) = A_1\exp(\frac{2}{k\rho_F^0}t)\Big((x+x_0) + \frac{1}{\chi}x_0^{1+\chi}(x+x_0)^{-\chi}\Big),$$

where $P_0$, $A_1$ and $x_0$ are arbitrary constants, $p_1 = -\frac{4x_0}{k^2\rho_F^0} < 0$, $\chi = -\frac{2\theta^1}{k\rho_F^0 D_1} < 0$, while other parameters are defined by properties of PEM and the solute transported throughout PEM.

3D plots of the exact solution (82) with specified parameters are presented in Fig.4–6 (a 3D plot for the function $\theta_F(t,x)$ is not shown because this component does not depend on time). These plots demonstrate that the above solution may describe (at least qualitative) the PEM layer deformation. As one concludes from the displacement plot in Fig.4, the largest deformation occurs in a vicinity of the end of the PEM layer $x = L$ where the external force (load) is acting. The function $u(t, x)$ is negative and this means that the layer is shrinking and the position of the point $x = L = 0.4$ in the time moment $t = 1$ is $L_S = 0.4 + (-0.2) = 0.2$, i.e. the width of the layer is two times smaller. The plot of the function $p(t, x)$ in Fig.5 shows that the hydrostatic pressure is growing with the time. It is natural because the layer is shrinking. The same behaviour (with respect to the time $t$) occurs also for the the concentration of molecules dissolved in fluid (see Fig.6).

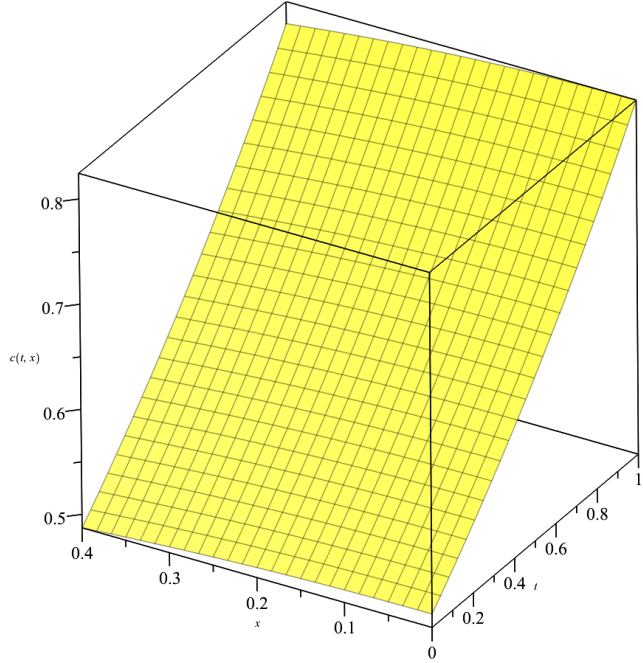

Figure 6: The function $c(t, x)$ represents the concentration of molecules dissolved in fluid.

## 6    Discussion

In this study, using mathematical foundations of the elasticity theory, a mathematical model for two solutes transport in a poroelastic material (soft tissue is a typical example) is suggested. It is assumed that molecules of essentially different sizes dissolved in fluid and are transported through pores of different sizes. The stress tensor, the main force leading to the material deformation, is taken not only in the standard linear form but also with an additional nonlinear part. The model is constructed in 1D space and consists of five linear and nonlinear equations (see formulae (27)–(31)).

It is shown that the governing equations of the model are integrable in stationary case, therefore all steady-state solutions are constructed. The obtained solutions are used in Example 1 for healthy and tumour tissue, in particular, tissue displacements are calculated for the parameters taken from experimental data (see Fig.2–3).

Since the governing equations are non-integrable in non-stationary case, the Lie symmetry method and standard methods for solving ODE systems are applied in order to construct time-dependent exact solutions. Depending on parameters arising in the governing equations, several special cases with non-trivial Lie symmetries are identified. Applying the symmetries obtained for reduction of the governing PDEs to systems of ODEs, multiparameter families of exact solutions are constructed including those in terms of special functions(hypergeometric and Bessel functions). A possible application of the solutions obtained is demonstrated in Example 2.

This work can be continued in different directions. A natural generalization of the model can be developed by taking into account the fact that there are real-world processes involving transportation of several solutes. A typical example can be the solute transport during peritoneal dialysis when a wide range of molecules of different sizes are removed or absorbed. In this case, the model should involve $N$ equations to find concentrations $c_1, c_2, \ldots, c_N$. We expect that Lie symmetry of the model obtained will be a direct generalization of that presented in Theorem 2. Thus, families of exact solutions, in particular steady-state solutions, can be derived in a similar way as it was done in this work.

Another generalization of the model can be obtained if one takes into account internal sources occurring in many processes including the peritoneal dialysis mentioned above. Such generalization will lead to more complicated equations and we predict that the Lie symmetry for system (33)—(39) will be not preserved. It means that the problem of construction of exact solutions can be much harder. An attempt in this direction was made in our recent paper [18].

We would like to point out also the following observation. Typically the Terzaghi effective stress tensor is used in the form

$$\tilde{\tau}_t = -p + (\lambda + 2\mu)u_x, \tag{83}$$

i.e. it is the linear function of the hydrostatic pressure and the dilatation $e = u_x$. Here we suggested that osmotic pressures and some nonlinearity w.r.t. the dilatation can contribute to the tensor. Moreover, the quadratic nonlinearity $\kappa u_x^2$ is only a simplest possibility. Generally speaking it can be a more complicated functions therefore the tensor can take the form

$$\tilde{\tau}_t = -p + (\gamma_0 + \gamma_1)RTc_1 + \gamma_2 RTc_2 + (\lambda + 2\mu)u_x + F(u_x), \tag{84}$$

where $F$ is some smooth function.

Finally, it should be noted that the work is in progress in order to generalize the model presented in this study on multidimensional cases, i.e. the 2D and 3D cases.

# 7 Acknowledgements

R.Ch. acknowledges that this research was funded by the National Center of Science of Poland (project UMO-2022/01/3/ST1/00097) and the British Academy (Leverhulme Researchers at Risk Research Support Grant).